\documentclass[]{elsart}
\usepackage[dvips]{graphicx}
\usepackage{array}
\usepackage{amssymb}
\usepackage{hhline}
\usepackage{longtable}
\usepackage{dcolumn}
\usepackage{bm}
\usepackage{subfigure}

\begin{document}

\begin{frontmatter}

\title{Effects of accelerating growth on the evolution of weighted complex networks}

\author[lable1,label2]{Zhongzhi Zhang}
\ead{zhangzz@fudan.edu.cn}
\author[lable1,label2]{Lujun Fang}
\ead{fanglujun@fudan.edu.cn}
\author[lable1,label2]{Shuigeng Zhou}
\ead{sgzhou@fudan.edu.cn}
\author[label3]{Jihong Guan}
\ead{jhguan@mail.tongji.edu.cn}

\address[lable1]{School of Computer Science, Fudan
University, Shanghai 200433, China}
\address[label2]{Shanghai Key Lab of Intelligent Information
Processing, Fudan University, Shanghai 200433, China}
\address[label3]{Department of Computer Science and Technology,
Tongji University, 4800 Cao'an Road, Shanghai 201804, China}

\begin{abstract}
Many real systems possess accelerating statistics where the total
number of edges grows faster than the network size. In this paper,
we propose a simple weighted network model with accelerating growth.
We derive analytical expressions for the evolutions and
distributions for strength, degree, and weight, which are relevant
to accelerating growth. We also find that accelerating growth
determines the clustering coefficient of the networks.
Interestingly, the distributions for strength, degree, and weight
display a transition from scale-free to exponential form when the
parameter with respect to accelerating growth increases from a small
to large value. All the theoretical predictions are successfully
contrasted with numerical simulations.
\end{abstract}

\end{frontmatter}

\maketitle
\section{Introduction}
Standard interesting objects in the network science are relatively
simple binary (Boolean) networks where edges (links) are either
present or absent, represented as binary
states~\cite{AlBa02,DoMe02,Ne03}. In other words, edges in Boolean
networks have equal ``weights". However, the connections in many
real networks are not homogeneous~\cite{BoLaMoChHw06}, which
naturally calls for a typical measurement of the edge weight, such
as the number of joint papers of two coauthors in scientific
collaboration
network~\cite{Ne01a,Newman01,BaJeNeRaScVi02,LiWuWaZhDiFa07}, the
number of flights or seats between any two cities in airline
networks~\cite{BaBaPaVe04,LiCa04}, the bandwidth of a link in the
Internet~\cite{FaFaFa99}, the reaction rate in metabolic
network~\cite{JeToAlOlBa00}, and so on. These real systems with
diversity of edges can be better described in terms of weighted
networks.

Various weighted network models have been proposed to describe and
explain the real-life systems~\cite{BoLaMoChHw06}. Yook \emph{et
al.} took a first step in the direction of a model study for
evolving weighted network (YJBT model)~\cite{YoJeBaTu01}, where the
topology and weight are driven by only the network connection based
on preferential attachment (PA) rule~\cite{BaAl99}. The YJBT model
overlooks the possible dynamical evolution and reinforcements of
weights, which is a common property of real-life
networks~\cite{BoLaMoChHw06}. To better mimic the reality, Barrat,
Barth\'elemy, and Vespignani presented a growing model (BBV) for
weighted networks, where the evolutions of degree and weight are
coupled~\cite{BaBaVe04a,BaBaVe04b}. Enlightened by BBV's remarkable
work, a variety of models and mechanisms for weighted networks have
been proposed, including weight-driven model~\cite{AnKr05},
traffic-driven evolution models~\cite{WaWaHuYaQu05,XiWaWa07},
fitness models~\cite{Bi05}, local-world
models~\cite{PaLiWa06,WaTaZhXi05,ZhRoWaZhGu07}, deterministic
models~\cite{DoMe05,ZhZhFaGuZh07,ZhZhChGuFaZh07}, weight-dependent
deactivation~\cite{WuXuWa05}, spatial
constraints~\cite{BaBaVe05,MuMa06}.

Recent empirical study demonstrated that many real natural and
social networks exhibit the characteristic of ``accelerating
growth", which means that the total number of edges increases more
quickly than linearly with the node (vertex) number. Generally,
networks with this property are called ``accelerating networks". For
instance, in both the World Wide Web~\cite{BrKuMaRaRaStToWi00} and
the Internet~\cite{FaFaFa99} the number of edges increases with time
in a nonlinear fashion. In metabolic network~\cite{JeToAlOlBa00},
the total number of links exceeds the total number of nodes by about
an order of magnitude. Other familiar examples of accelerating
networks include scientific collaboration
network~\cite{Ne01a,Newman01,BaJeNeRaScVi02,LiWuWaZhDiFa07},
language network~\cite{DoMe01a}, citation network~\cite {Re98}, etc.
Inspired by this phenomenon, the research on accelerating networks
has attracted an amount of
attention~\cite{DoMe01b,Se04,MaGa05,GaMa05,YuZhLiWaXu06}.

Actually, accelerating networks are far more common in the real
world than has hitherto been appreciated~\cite{MaGa05}. Despite
their widespread appearance, evolving weighted networks with
accelerating growth have received less attention. This important
factor of accelerating growth is neglected in most of the considered
weighted network
models~\cite{YoJeBaTu01,BaBaVe04a,BaBaVe04b,AnKr05,WaWaHuYaQu05,XiWaWa07,Bi05,PaLiWa06,DoMe05,ZhZhFaGuZh07,ZhZhChGuFaZh07,WuXuWa05,BaBaVe05,MuMa06}.
Then questions arise naturally: Can an accelerating growth model for
weighted networks be established? How much effect does the
accelerating growth have on the evolution of weighted networks?

In this paper, we present an accelerating weighted network model to
understand how the accelerating growth phenomenon affects the
dynamical evolution of weighted networks. We study both analytically
and numerically the network characteristics, including the evolution
and distributions of the degree, weight, and strength, as well as
the clustering coefficient. We show that obtained properties depend
on the accelerating parameter.

\section{Definitions}
We give a brief introduction to the definitions of tools for
statistical characterization of weighted networks.

Mathematically, for a weighted network, its topological as well as
weighted properties can be completely described by a generalized
adjacency matrix $W$, whose element $w_{ij}$ specifies the weight of
the edge between node $i$ and $j$. $w_{ij}=0$ represents that node
$i$ and $j$ are disconnected. In the following, we focus on the
cases of undirected graphs, which have symmetric nonnegative weights
$w_{ij}=w_{ji}\geq 0$. Moreover, we assure that $w_{ii}=0$.

The standard topological characterization of binary networks is also
applied for weighted networks, which is obtained by the analysis of
the distribution $P(k)$ that represents the probability of a random
selected node to have degree $k$. In a weighted network, a natural
generalization of degree is the node strength defined as
$s_{i}=\sum_{j\in \Gamma (i)} w_{ij}$, where the sum runs over the
set $\Gamma (i)$ of neighbors of the node $i$. Statistical
properties of weighted networks can be characterized by the
distributions of strength $P(s)$ and weight $P(w)$, which denote the
probability of a node to have strength $s$ and of an edge to have
weight $w$.

\section{The model for accelerating weighted networks}
The model proposed here begins from an initial configuration of
$N_{0}$ nodes connected by edges with assigned weight $w_{0}=1$. At
each time step, the network evolves under the following two coupled
mechanisms: topological growth and weights' dynamics.

(i) Topological growth. A new node $n$ enters the network. If the
new node is born at time $t$, we assume that the edge number of this
new node is a power law function $t^{\theta}$ ($0\leq \theta \leq1$)
depending on time $t$, where $\theta$ is the acceleration parameter.
(Since multiple edges are forbidden, the total number of edges is
smaller than $t^2/2$, thus $\theta$ cannot be greater than 1. On the
other hand, when $\theta <0$, the new node may carry no edge, which
is not consistent with most real networks. Thus, one may reasonably
assume $0\leq \theta \leq1$.) These $t^{\theta}$ new edges have
initial weight $w_{0}=1$ and are randomly attached to a previously
existing node $i$ according to the preferential probability
\begin{equation}\label{pi}
\Pi_{n\rightarrow i}=\frac{s_{i}}{\sum_{j}s_{j}} \,,
\end{equation}
where the sum runs over all existing nodes.

(ii) Weights' dynamics. The creation of each of the $t^{\theta}$
edges will introduce variations of the existing weights across the
network. For the sake of the simplicity, we only consider the local
rearrangements of weights of those edges connecting $i$ and its
neighbors $j\in \Gamma (i)$, according to the simple rule
\begin{equation}\label{wij}
w_{ij}\rightarrow w_{ij}+ \delta \frac{w_{ij}}{s_{i}} \,.
\end{equation}
Here we have assumed the addition of each new edge induces a total
increment $\delta$ ($\delta$=const) of weights. The rule described
by Eq.~(\ref{wij}) yields a global increase of $w_{0}+\delta$ for
the strength of node $i$, which will therefore become even more
attractive to future nodes.

After updating the weights, the growing process is iterated by
introducing another new node, i.e. returning to step (i) until the
network reaches the desired size. Since the network size is
incremented by one with each time step, we use the step value $t$ to
represent the node created at this step. At time $t$, the network
has $N=t+N_{0}$ nodes and  $\int t^{\theta} d t =t^{1+\theta}/
(1+\theta)$ edges in the continuum limit.

Note that many real-life networks exhibit such an evolving mechanism
as described in our model. Mechanism (i) is a plausible one that
appears in many real systems such as the Internet and scientific
collaboration networks; it corresponds to the fact that new nodes
try to connect a preexisting node with a probability proportional to
the strength of the old node, as provided by Eq.~(\ref{pi}). An
important aspect of mechanism (i) is that the number of edges
carried by the new node is not a constant, but controlled by a
parameter $\theta$ $(0\leq \theta \leq1)$, which we call
accelerating exponent. This has been confirmed by a variety of
empirical observations. For example, for arXiv citation graph,
autonomous system graph of the Internet, and the email networks,
their accelerating exponents have been found to be 0.56, 0.18, and
0.12, respectively~\cite{LeKlFa07}. On the other hand, mechanism
(ii) describes the weight dynamics induced by a new edge onto the
old ones, which can happen in scientific collaboration networks,
airline networks, and so on. See
Refs.~\cite{BaBaPaVe04,BaBaVe04a,BaBaVe04b} for detailed
explanation.

The model is governed by two parameters $\delta$ and $\theta$,
according to which there are some limiting cases of the model. When
$\delta=0$ and $\theta=0$, it is reduced to the BA
model~\cite{BaAl99}. For $\theta=0$, it coincides with a special
case of the BBV model~\cite{BaBaVe04a,BaBaVe04b}. Varying $\delta$
and $\theta$ allows one to study the crossover between the two
limiting models, which have qualitatively different properties from
the two limits. We will show that both of the parameters $\delta$
and $\theta$ have significant effects on the network evolution, here
we focus on the latter that has never been studied before, while the
former has been discussed in Ref.~\cite{BaBaVe04b}.

\section{Evolution and distributions of strength, degree and weight}

Our growing model can be studied analytically through the time
evolution of the average value of $s_{i}(t)$ and $k_{i}(t)$ of the
$i$th node at time $t$ by using the ``mean-field'' method. According
to the evolution rules, the addition of each new edge results in the
increase of total network strength by an amount equal to
$2+2\delta$. Thus, after $t$ steps of evolution, the total strength
of the network is obtained to be
\begin{eqnarray}\label{s}
\sum_{i}s_{i}(t)&= \sum_{i=0}^{t}i^{\theta}(2+2\delta)\nonumber\\
& \approx \int_{0}^{t}i^{\theta}(2+2\delta)di\nonumber\\
&=\frac{2(1+\delta)}{\theta+1}t^{\theta+1}\,.
\end{eqnarray}

When a new node $n$ enters the network, an existing node $i$ can be
affected in two ways: (1) The new node is connected to $i$ with
probability given by Eq. (\ref{pi}), thus the degree and strength of
$i$ are increased by 1 and $1+\delta$, respectively. (2) The new
node is connected to one of $i$'s neighbors $j\in \Gamma (i)$, in
this case the degree of $i$ remains unmodified, while $w_{ij}$ is
increased according to Eq. (\ref{wij}), and thus $s_{i}$ is
increased by $\delta w_{ij}/s_{j}$. We assume that variables $s_{i}$
and $k_{i}$ are continuous. Then, at each time step, the strength
$s_{i}$ and degree $k_{i}$ of a node evolve as
\begin{eqnarray}\label{si01}
\frac{ds_{i}}{dt}&= t^{\theta}\frac{s_{i}}{\sum_{j}s_{j}}(1+\delta) + \sum_{j\in \Gamma (i)}t^{\theta}\frac{s_{j}}{\sum_{k}s_{k}}\delta\frac{w_{ij}}{s_{j}}\nonumber\\
 &= \frac{2\delta+1}{2\delta+2}(1+\theta)\frac{s_{i}(t)}{t}\,,
\end{eqnarray}
and
\begin{equation}\label{ki01}
\frac{dk_{i}}{dt} =
t^{\theta}\frac{s_{i}(t)}{\sum_{j}s_{j}(t)}=\frac{1+\theta}{2+2\delta}\frac{s_{i}(t)}{t},
\end{equation}
respectively.

\begin{figure}
\begin{center}
\includegraphics[width=0.55\textwidth]{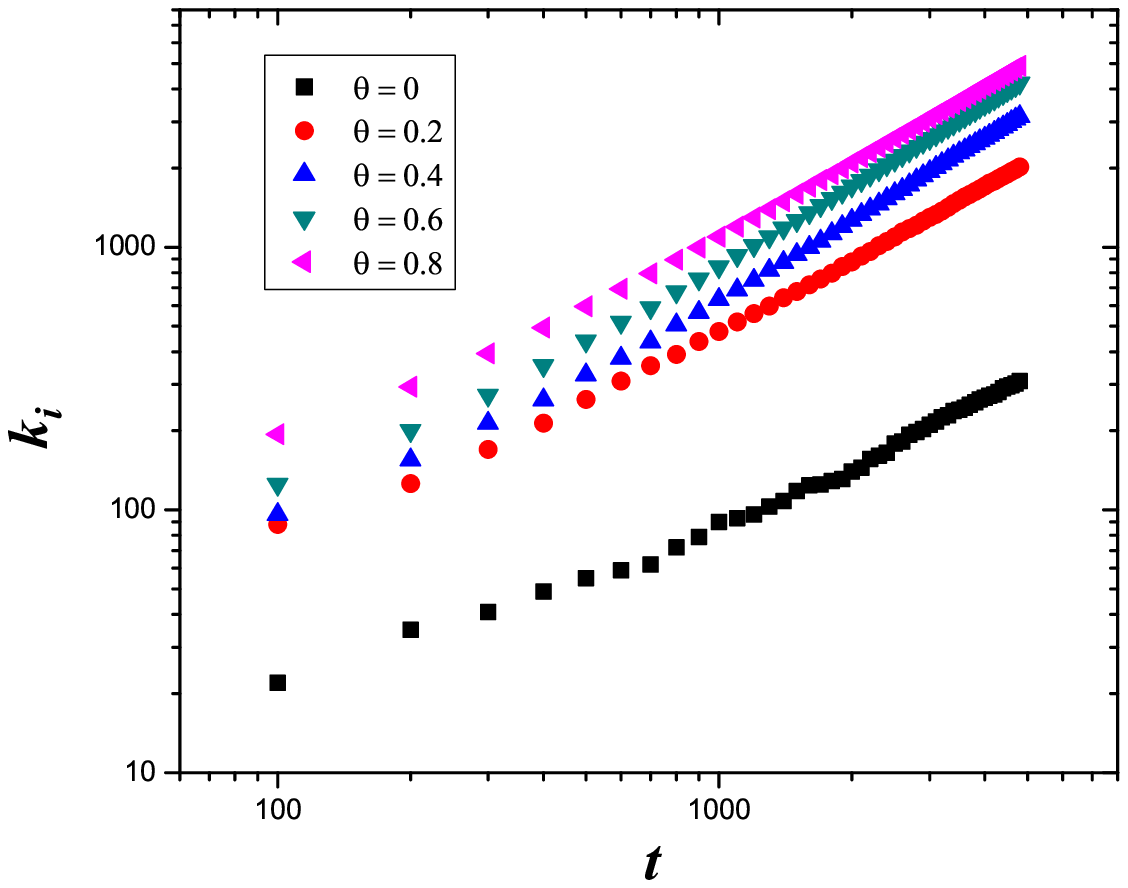}
\includegraphics[width=0.55\textwidth]{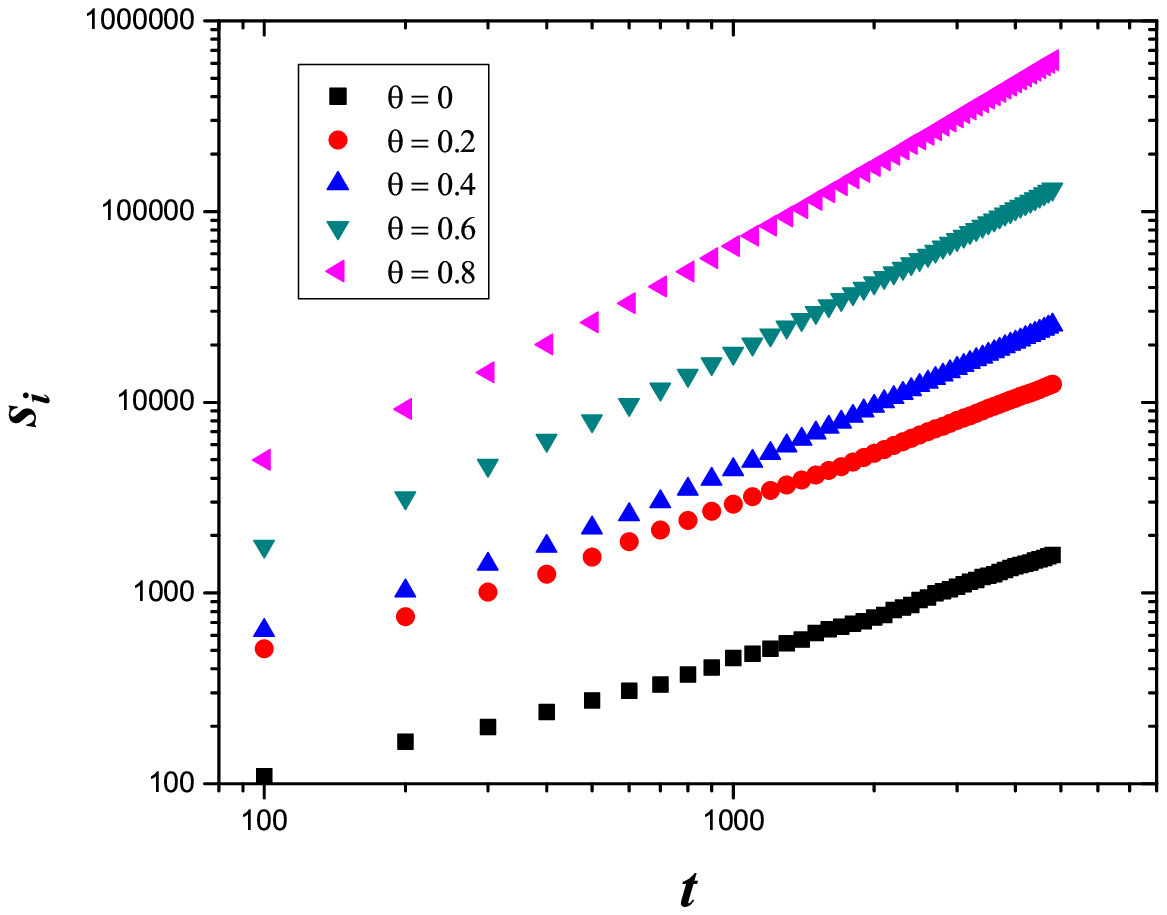}
\caption[kurzform]{\label{Evolution} (Color online) Time-evolution
for the degree and strength of nodes, added to the systems started
at $t= 1$ in the case of $\delta=2$.}
\end{center}
\end{figure}

With the initial conditions $k_{i}(t=i)=s_{i}(t=i)=i^{\theta}$, we
can integrate above equations to obtain
\begin{equation}\label{si02}
s_{i}(t) =i^{\theta}\left(\frac{t}{i}\right)^{\beta}=
i^{\theta}\left(\frac{t}{i}\right)^{\frac{(2\delta+1)(1+\theta)}{2\delta+2}},
\end{equation}
and
\begin{eqnarray}\label{ki02}
k_{i}(t) &= \frac{s_{i}(t)+2\delta t^{\theta}}{2\delta+1}
\nonumber\\
&=\frac{i^{\theta}\left(\frac{t}{i}\right)^{\frac{(2\delta+1)(1+\theta)}{2\delta+2}}+2\delta
t^{\theta}}{2\delta+1}.
\end{eqnarray}
For large $t$, it can be always guaranteed that $s_{i}(t)$ is larger
than $2\delta t^{\theta}$ because one can easily see that
$\frac{(2\delta+1)(1+\theta)}{2\delta+2}>\theta$ given $\theta\in
[0, 1]$ and $\delta>0$. Thus, $s_{i}(t)$ and $ k_{i}(t)$ are
proportional when $t$ is large. From the obtained expressions, we
find that dynamical exponent
$\beta=\frac{(2\delta+1)(1+\theta)}{2\delta+2}$ depends on not only
$\delta$, but also $\theta$, which is significantly different from
the BBV model and BA network. Figure \ref{Evolution} shows the
behavior of the nodes' degree and strength versus time for different
$\theta$, which recovers the results predicted analytically.

We assume that the nodes are added to the systems at equal time
intervals, then the probability that a node has strength $s_{i}(t)$
smaller than  $s$, $P(s_{i}(t)<s)$, can be written as
\begin{eqnarray}
P\left(s_{i}(t)<s\right) &=&
P\left(i>s^{-\frac{2\delta+2}{2\delta+1-\theta}}t^{\frac{(2\delta+1)(1+\theta)}{2\delta+1-\theta}}\right)\nonumber\\
&=&1-P\left(i\leq s^{-\frac{2\delta+2}{2\delta+1-\theta}}t^{\frac{(2\delta+1)(1+\theta)}{2\delta+1-\theta}}\right)\nonumber\\
&=&1-\frac{1}{N_{0}+t}s^{-\frac{2\delta+2}{2\delta+1-\theta}}t^{\frac{(2\delta+1)(1+\theta)}{2\delta+1-\theta}}.
\end{eqnarray}
The probability distribution of strength $P(s)$ can be calculated
through solving the partial differentiation of $P(s_{i}(t)<s)$ on
$s$, and the final result of strength distribution at time $t$
exhibits the following behavior:
\begin{equation}\label{ps}
P(s)=\frac{2\delta+2}{2\delta+1-\theta}f(t)s^{-\frac{4\delta+3-\theta}{2\delta+1-\theta}},
\end{equation}
where
$f(t)=\frac{1}{t+N_{0}}\,t^{(2\delta+1)(1+\theta)/(2\delta+1-\theta)}$.
Therefore, the strength follows a power law distribution with an
 exponent
\begin{equation}\label{gammas}
\gamma_{s} = \frac{4\delta+3-\theta}{2\delta+1-\theta}\,.
\end{equation}

Since there is an approximatively linear relationship between the
degree and strength for the same node, the degree distribution
$P(k)$ has also a scale-free form,  $P(k)\sim k^{-\gamma_{k}}$
 with the exponent $\gamma_{k}$ identical to $\gamma_{s}$.

\begin{figure}
\begin{center}
\includegraphics[width=0.45\textwidth]{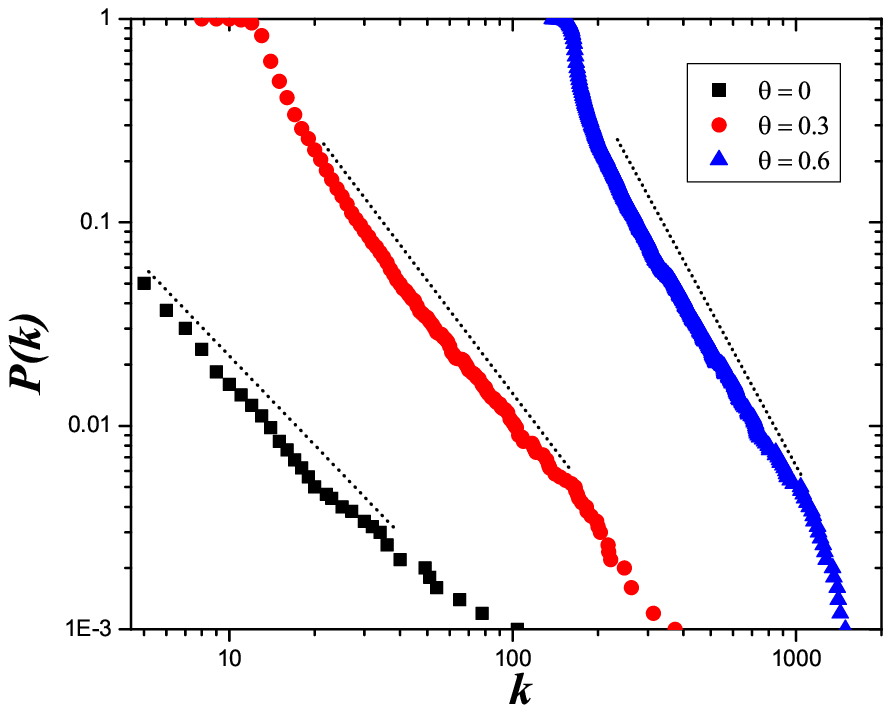}\\
\includegraphics[width=0.45\textwidth]{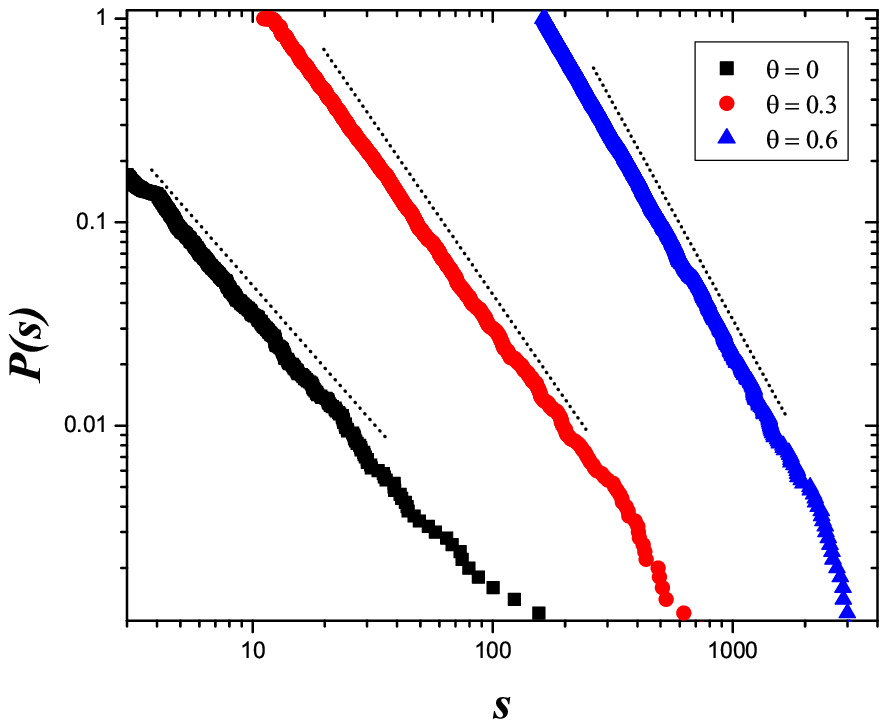}
\caption[kurzform]{\label{NodeDistribution} (Color online) The
cumulative distributions of nodes'  degree and strength at various
$\theta$ values for $\delta=5$. The network size is $N=5000$, and
the straight lines are the theoretical results of
$\gamma_{k,cum}=\gamma_{s,cum}=(4\delta+3-\theta)/(2\delta+1-\theta)-1$.}
\end{center}
\end{figure}

In order to confirm the validity of the obtained analytical
predictions, we performed extensive numerical simulations of the
networks. To reduce the effect of fluctuation on simulation results,
the simulation results are average over ten network realizations.
Figure~\ref{NodeDistribution} gives the accumulative distributions
of nodes' degree and strength. Numerical results are consistent with
the theoretical ones.

Now we investigate the evolution of the weights $w_{ij}$ with time,
which can also be computed analytically using mean field
approximation employed for the research of $s_{i}(t)$ and
$k_{i}(t)$. During the process of network growth, $w_{ij}$ can only
increase by the addition of a new node connected to either node $ i$
or $j$, and the evolution of  $w_{ij}$ satisfies the following
equation
\begin{eqnarray}\label{wij01}
\frac{dw_{ij}}{dt}&=t^{\theta}\frac{s_{i}}{\sum_{k}s_{k}}\delta\frac{w_{ij}}{s_{i}}
+t^{\theta}\frac{s_{j}}{\sum_{k}s_{k}}\delta\frac{w_{ij}}{s_{j}}\nonumber\\
&=\frac{\delta(1+\theta)}{1+\delta}\frac{w_{ij}}{t}.
\end{eqnarray}

\begin{figure}
\begin{center}
\includegraphics[width=0.55\textwidth]{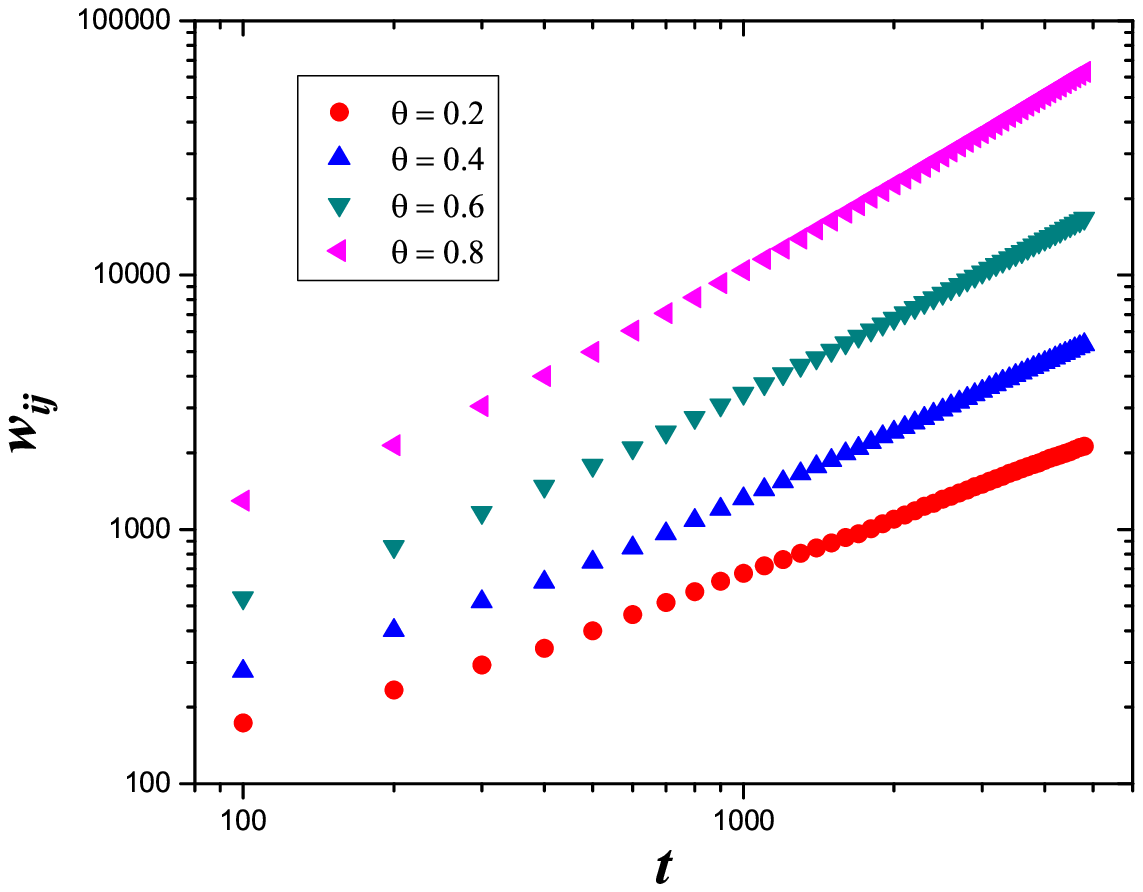}
\includegraphics[width=0.46\textwidth]{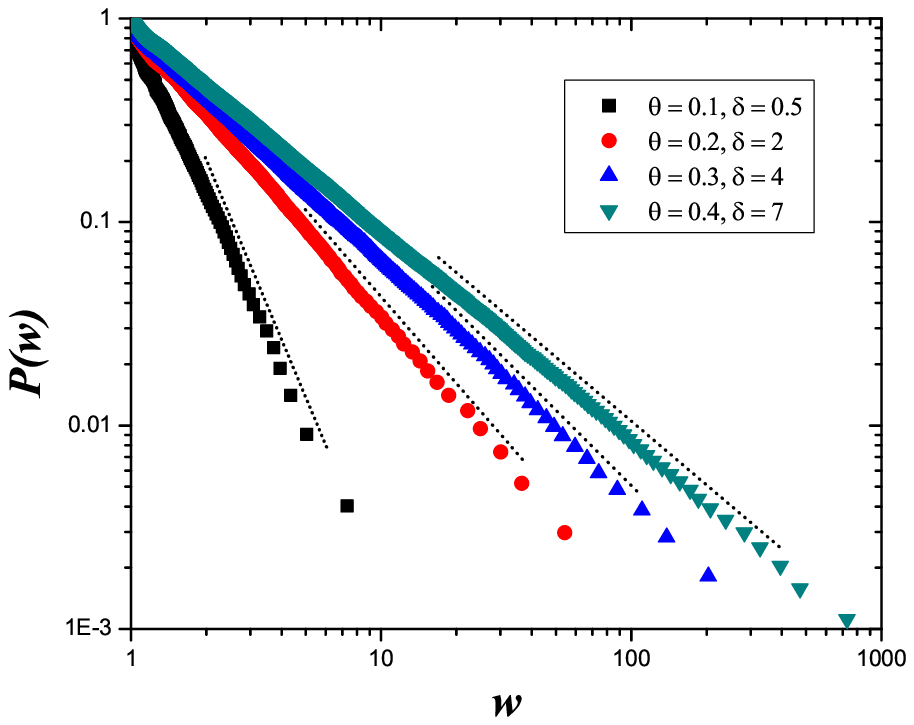}
\caption[kurzform]{\label{WeightDistribution} (Color online) (Upper
panel) Log-log plot of time-evolution for the weight of edges, which
are born at step $t= 1$ in the case of $\delta=2$. (Lower panel) The
cumulative distribution of edge weight for network size $N=5000$ at
different $\theta$ and $\delta$. The dashed straight lines are the
theoretical results of $\gamma_{w,cum}=(1+2\delta+\delta
\theta)/[(1+\theta)\delta]-1$.}
\end{center}
\end{figure}

The edge $(i,j)$ is built only when both node $i$ and $j$ have been
created, therefore the birth time of edge $(i,j)$ is $t_{ij} =
max(i,j)$. Considering the initial condition $w_{ij}=1$, one can
integrate the Eq.~(\ref{wij01}) to obtain
\begin{equation}\label{wij02}
w_{ij}(t)=\left(\frac{t}{t_{ij}}\right)^{\frac{\delta(1+\theta)}{\delta+1}},
\end{equation}
which implies that the edge weight $w$ displays a power-law
distribution, $P(w)\sim w^{-\gamma_{w}}$, with the exponent
$\gamma_{w}=\frac{1+2\delta+\delta \theta}{(1+\theta)\delta}$. To
check the validity of the analytical predictions for the evolution
and distribution of the edge weight, we have performed numerical
simulations of the present model, which are plotted in
Fig.~\ref{WeightDistribution}. The simulations are in agreement with
the analytical calculations.

So far, we have shown that the considered accelerating weighted
network has power-law distributions of strength, degree, and weight.
All the obtained exponents
$\gamma_{k}=\gamma_{s}=\frac{4\delta+3-\theta}{2\delta+1-\theta}$,
and $\gamma_{w}=\frac{1+2\delta+\delta \theta}{(1+\theta)\delta}$
vary from 2 to $\infty$, depending on the network parameters
$\delta$ and $\theta$. It should be noted that in the BBV model,
where $\theta$ equals zero, exponents $\gamma_{k}=\gamma_{s}$ are
between 2 and 3. Thus, accelerating growth has an important effect
on the evolution of the network: when $\theta$ increases from 0 to
1, distributions of strength, degree, and weight exhibit a
transition from scale-free (small exponent) to exponential (large
exponent) forms. Moreover, by tuning the values of parameter
$\theta$, the network may have different forms of degree (strength)
distribution and weight distribution. For example, when $\delta \to
0$, the exponents $\gamma_{k}$ ($\gamma_{s}$) and $\gamma_{w}$ are
reduced to $\gamma_{k}=\gamma_{s}=1+\frac{2}{1-\theta}$ and
$\gamma_{w}=\infty$, respectively. In this case, the weight
distribution is always exponential, while the degree (strength)
distribution follows a power-law form with $\gamma_{k}$
($\gamma_{s}$) increasing from 3 to $\infty$ when parameter $\theta$
increases from 0 to 1.

\section{Clustering coefficient}
As studied in the previous section, accelerating growth
significantly affects the network properties, such as distributions
of strength, degree, and weight. In this section, we will show that
the parameter $\theta$ concerning  accelerating growth also controls
the clustering coefficient of the networks.

By definition, the clustering coefficient~\cite{WaSt98} of a given
node is the ratio of the total number of edges that actually exist
between all its $k$ nearest neighbors and the potential number of
edges $k(k-1)/2$ between them. The clustering coefficient of the
whole network is obtained through averaging over all its nodes.

\begin{figure}
\begin{center}
\includegraphics[width=0.55\textwidth]{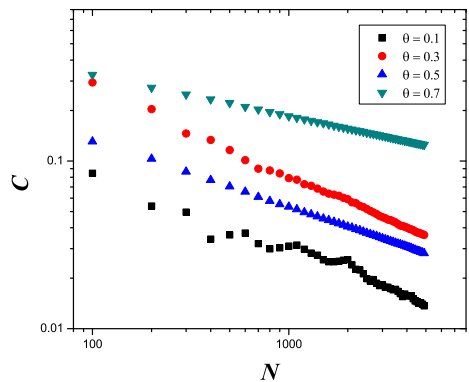}
\caption[kurzform]{\label{clustering} (Color online) Average
clustering coefficient $C$ vs parameter $\theta$ and  the network
size $N$. All data are from the average of ten independent
simulations for the same network.}
\end{center}
\end{figure}
We have performed numerical simulations of the networks to study the
influence of the acceleration parameter $\theta$ on clustering
coefficient, which is presented in Fig.~\ref{clustering}. Numerical
simulation results show that for arbitrary $\theta$, the average
clustering coefficient $C(N)$ decreases as the size increases, i.e.
$C(N)\sim N^{-\eta}$, as reported in Fig.~\ref{clustering}. This
power-law behavior is similar to that of Barab\'asi-Albert (BA)
model~\cite{AlBa02}, and is in contrast to the observation of some
real networks such as the Internet and metabolic networks whose
clustering coefficient is independent of their size~\cite{RaBa03}.
However, to the best of our knowledge, whether there is real
networks exhibiting similar phenomenon of clustering coefficient as
our model is still unknown, which deserves further study in future.

In fact, there are two known limits for the average clustering
coefficient $C$ of the whole network: In the case of $\theta=0$, the
network is a tree, and hence it has no triangle, therefore $C=0$;
when $\theta=1$, the network corresponds to a complete graph (i.e.
$N$-clique) with clustering coefficient $C=1$.
Figure~\ref{clustering} also shows the dependence of $C$ on
$\theta$. One can easily see that $C$ is an increasing function of
$\theta$ as expected. The increase is not very sharp for small
$\theta$, but we can expect an obvious increase of $C$ for $\theta$
of very large values.

\section{Conclusion} 

In conclusion, we have proposed a growing model for weighted
networks in which the number of edges added with each new node is an
increasing function (power law function) of network size. We have
demonstrated that the accelerating growth is an important factor
that establishes the network structure. Using mean field network
theory, we have computed analytical expressions for the evolution
and distributions of strength, degree and weight. The obtained
results show that these distributions are subject to a transition
from a power-law to exponential shape, when the acceleration
parameter is tuned from a small to large value.  All mean field
approximations have been confirmed by numerical simulations. Our
model may provide valuable insight into the real-life networks.

Although accelerating growth exists in many real-life networks, it
should be pointed out that the nonlinear growth fashion of the edges
compared with nodes in real systems is more intricate and flexible.
We use here the most generic case, i.e. the number of total edges is
a power-law function of the total node number. In future, it would
be worth studying in detail other manners of nonlinear growth in
different real-life networks as well as their impacts on network
properties and dynamics.

\section*{Acknowledgments}
The authors are grateful to the anonymous referees for their
valuable comments and suggestions. This research was supported by
the National Basic Research Program of China under grant No.
2007CB310806, the National Natural Science Foundation of China under
Grant Nos. 60496327, 60573183, 60773123, and 60704044, the Shanghai
Natural Science Foundation under Grant No. 06ZR14013, the
Postdoctoral Science Foundation of China under Grant No.
20060400162, Shanghai Leading Academic Discipline Project No. B114,
the Program for New Century Excellent Talents in University of China
(NCET-06-0376), and the Huawei Foundation of Science and Technology
(YJCB2007031IN).
%

\begin{thebibliography}{10}
\bibitem{AlBa02} R. Albert and A.-L. Barab\'asi,
       Rev. Mod. Phys. {\bf 74}, 47 (2002).

\bibitem{DoMe02} S. N. Dorogvtsev and J.F.F. Mendes,
Adv. Phys. {\bf 51}, 1079 (2002).

\bibitem{Ne03} M. E. J. Newman,
SIAM Review {\bf 45}, 167 (2003).


\bibitem{BoLaMoChHw06}
S. Boccaletti, V. Latora, Y. Moreno, M. Chavezf, and D.-U. Hwanga,
Phy. Rep. {\bf 424}, 175 (2006).


\bibitem{Ne01a}
M. E. J. Newman, Proc. Natl. Acad. Sci. U.S.A. {\bf 98}, 404 (2001).

\bibitem{Newman01}
M. E.~J. Newman, Phys. Rev. E {\bf 64}, 016132 (2001).

\bibitem{BaJeNeRaScVi02}
A.-L. Barab\'asi, H. Jeong, Z. N\'eda. E. Ravasz, A. Schubert, and
T. Vicsek, Physica A {\bf 311}, 590 (2002).

\bibitem{LiWuWaZhDiFa07}
M. Li, J. Wu, D. Wang, T. Zhou, Z. Di, Y. Fan, Physica A {\bf 375},
355 (2007).

\bibitem{BaBaPaVe04}
A. Barrat, M. Barth\'elemy, R. Pastor-Satorras, and A. Vespignani,
Proc. Natl. Acad. Sci. U.S.A. {\bf 101}, 3747 (2004).

\bibitem{LiCa04}
W. Li, and X. Cai, Phys. Rev. E {\bf 69}, 046106 (2004).

\bibitem{FaFaFa99}
M. Faloutsos, P. Faloutsos and C. Faloutsos, Comput. Commun. Rev.
{\bf 29}, 251 (1999).


\bibitem{JeToAlOlBa00}
H. Jeong, B. Tombor, R. Albert, Z. N. Oltvai and A.-L. Barab\'asi,
Nature {\bf 407}, 651 (2000).


\bibitem{YoJeBaTu01}
S. H. Yook, H. Jeong, A.-L. Barab\'asi, Y. Tu, Phys. Rev. Lett. {\bf
86}, 5835 (2001).


\bibitem{BaAl99} A.-L. Barab\'asi and R. Albert,
       Science {\bf 286}, 509 (1999).


\bibitem{BaBaVe04a} A. Barrat, M. Barth\'elemy, and A. Vespignani, Phys.
Rev. Lett. {\bf 92}, 228701 (2004).

\bibitem{BaBaVe04b} A. Barrat, M. Barth\'elemy, and A. Vespignani, Phys.
Rev. E {\bf 70}, 066149 (2004).

\bibitem{AnKr05}
T. Antal and P. L. Krapivsky, Phys. Rev. E {\bf 71} 026103 (2005).

\bibitem{WaWaHuYaQu05}
W.-X. Wang, B.-H. Wang, B. Hu, G. Yan, and Q. Ou, Phys. Rev. Lett.
\textbf{94}, 188702 (2005).

\bibitem{XiWaWa07}
Y.-B. Xie, W.-X. Wang, and B.-H. Wang, Phys. Rev. E \textbf{75},
026111 (2007).

\bibitem{Bi05}
G. Bianconi, Europhys. Lett. \textbf{71}, 1029 (2005).

\bibitem{PaLiWa06}
Z. Pan, X. Li, and X. Wang, Phys. Rev. E \textbf{73}, 056109 (2006).

\bibitem{WaTaZhXi05}
B. Wang, H. W. Tang, Z. Z. Zhang, and Z. L. Xiu,
Int. J. Mod. Phys. B {\bf 19}, 3951 (2005).

\bibitem{ZhRoWaZhGu07}
Z.Z. Zhang, L. L. Rong, B. Wang, S. G. Zhou, and J. H. Guan, Physica
A {\bf 380}, 639 (2007).

\bibitem{DoMe05} S. N. Dorogvtsev and J. F. F. Mendes,
AIP Conf. Proc. {\bf 776}, 29 (2005).

\bibitem{ZhZhFaGuZh07}
Z.Z. Zhang, S.G. Zhou, L.J. Fang, J.H. Guan,  Y.C. Zhang, EPL {\bf
79}, 38007 (2007)

\bibitem{ZhZhChGuFaZh07}
Z.Z. Zhang, S. G. Zhou, L. C. Chen, J. H. Guan, L. J. Fang, and Y.
C. Zhang, Eur. Phys. J. B {\bf 59}, 99 (2007).

\bibitem{WuXuWa05}
Z.-X. Wu, X.-J. Xu, and Y.-H. Wang, Phys. Rev. E {\bf 71}, 066124
(2005).

\bibitem{BaBaVe05} A. Barrat, M. Barth\'elemy, and A. Vespignani, J. Stat. Mech: Theory Exp. {\bf P05003} (2005).

\bibitem{MuMa06}
G. Mukherjee and S. S. Manna, Phys. Rev. E {\bf 74}, 036111 (2006).

\bibitem{BrKuMaRaRaStToWi00}
A. Broder, R. Kumar, F. Maghoul, P. Raghavan, S. Rajagopalan, R.
Stata, A. Tomkins, and J. Wiener, Comput. Netw. {\bf 33}, 309
(2000).

\bibitem{DoMe01a}
S. N. Dorogovtsev and J. F. F. Mendes, Proc. R. Soc. London, Ser. B
{\bf 268}, 2261 (2001)


\bibitem{Re98}
S. Redner, Eur. Phys. J. B {\bf 4}, 131 (1998).

\bibitem{DoMe01b}
S. N. Dorogovtsev and J. F. F. Mendes, Phys. Rev. E {\bf 63},
025101(R) (2001).

\bibitem{Se04}
P. Sen, Phys. Rev. E {\bf 69}, 046107 (2004).

\bibitem{MaGa05}
J. S. Mattick and M. J. Gagen, Science {\bf 307}, 856 (2005).

\bibitem{GaMa05}
M. J. Gagen and J. S. Mattick, Phys. Rev. E {\bf 72}, 016123 (2005).

\bibitem{YuZhLiWaXu06}
X. Yu, Z. Li, D. Zhang, F. Liang, X. Wang, and X. Wu, J. Phys. A
{\bf 39}, 14343 (2006).

\bibitem{LeKlFa07}
J. Leskovec, J. Kleinberg, C. Faloutsos, ACM Transactions on
Knowledge Discovery from Data, {\bf 1}, 1 (2007).



\bibitem{WaSt98} D. J. Watts and H. Strogatz,
        Nature (London) {\bf 393}, 440 (1998).

\bibitem{RaBa03}
E. Ravasz, A.-L. Barab\'asi, Phys. Rev. E 67, 026112 (2003).


\end{thebibliography}
%

\end{document}